\documentclass[aps,twocolumn]{revtex4}
\usepackage{epic}
\usepackage{eepic}
\newcommand{\be}{\begin{equation}}
\newcommand{\ee}{\end{equation}}
\newcommand{\bea}{\begin{eqnarray}}
\newcommand{\eea}{\end{eqnarray}}

\newcommand{\al}{\alpha}
\newcommand{\bt}{\beta}
\newcommand{\ta}{\tau}
\newcommand{\s}{\sigma}
\newcommand{\g}{\gamma}

\newcommand{\hs}[1]{\hspace{#1 mm}}

\begin{document}

\title{Open Cosmic Strings in Black Hole Space-Times} 
\author{Ali Kaya} 
\email[e-mail: ]{kaya@gursey.gov.tr}
\affiliation{Feza Gursey Institute\\ Emek Mah. No:68,
Cengelkoy 81220 \\ Istanbul, Turkey}

\begin{abstract}
We construct open cosmic string solutions in Schwarzschild black
hole and non-dilatonic black $p$-brane backgrounds. 
These strings can be thought to stretch between two D-branes or
between a D-brane and the horizon in curved space-time. We study small
fluctuations around these solutions and discuss their basic
properties.
\end{abstract}


\maketitle

\section{Introduction}

General relativity offers a very  good description of gravity in terms
of a  curved space-time metric. However, this  description breaks down
when  a sufficiently  large distribution  of matter  collapses  due to
gravitational attraction. At the end of the collapse, a singularity is
inevitable provided  that general relativity  is correct and  that the
strong  energy condition  holds for  matter. According  to  the cosmic
censor conjecture, no singularity is ever visible to any observer. The
endpoint of a  collapse will be a black hole  where the singularity is
located  in a  region  of space-time  causally  disconnected from  the
asymptotic observer.   Even if the  cosmic censor conjecture  is true,
one may still  insist on a well defined  physical picture, rather than
having a secret  region where physics is no  longer operational. It is
natural to  expect that quantum gravitational effects  to be important
in  the very  strong fields  near  a singularity.  More precisely,  we
expect  that  at  the   Planck  scale  the  classical  description  of
space-time becomes completely inadequate.

\
\

Classically, nothing  can escape from a black  hole.  However, Hawking
showed  that  black holes  radiate  energy  as  if they  were  perfect
blackbody  emitters   with  the  temperature   $T=\kappa/2\pi$,  where
$\kappa$     is    the    surface     gravity    at     the    horizon
\cite{h1}. Qualitatively, the radiation can be thought to arise due to
vacuum fluctuations  of matter fields  near the horizon. The  pairs of
particles  and anti-particles  are produced,  and as  one  member with
negative energy  falls into the hole,  the other member  can escape to
infinity. Moreover, as the  black hole emits radiation its temperature
increases. Therefore,  one expects that  a black hole radiates  all of
it's  energy in  a finite  proper time  and disappears.   In Hawking's
derivation, it  turns out that the  radiation coming from  the hole is
insensitive to the  details of matter that makes up  the hole.  If the
semiclassical reasoning of Hawking is completely reliable at all times
during the  evaporation, then  this implies an  important modification
for  quantum  mechanics  applied  to general  relativity.  Namely,  in
quantum gravity it is possible to have a process where an initial pure
state  (matter  before the  collapse)  may  evolve  to a  mixed  state
(thermal radiation) \cite{h2}.

\
\

To avoid such a possibility,  one should assume that Hawking radiation
can carry out information about the collapsed matter.  Since the pairs
are created  near the  horizon, one would  suggest to  include quantum
gravitational  effects in  describing near  horizon  physics. However,
curvature invariants near the horizon  scales as $1/M^2$, where $M$ is
the mass  of the hole.  Thus  for a sufficiently heavy  black hole the
near horizon geometry  is locally close to flat  space.  Moreover, the
distance   between  the   collapsed  matter   concentrated   near  the
singularity  and  the horizon  is  proportional  to  $M$.  Thus,  this
distance  can in  principle be  much greater  than the  Planck length,
which  is  the  scale   where  quantum  gravitational  effects  become
important.  Therefore,  it seems that the problem  should be addressed
and solved in a semiclassical context.

\
\

Recent  progress in string  theory strongly  suggests that  black hole
evaporation is a unitary process. However, it is hard (if it is indeed
possible)  to   quantize  string  theory  exactly  in   a  black  hole
background.   In this  situation,  one can  implement a  semiclassical
approximation  near a  suitable  classical solution.   As pointed  out
above,  such an  approximation  might  also offer  a  solution to  the
information  puzzle.  A  number of  {\it closed}  string  solutions in
classical  backgrounds have  been found  and studied  before  (see for
instance \cite{c0}-\cite{c7}).  For  example, in \cite{c7}, the center
of mass  trajectories were classified in Schwarzschild  black hole and
semi-classical  quantization  were  studied  to  the  first  order  in
fluctuations.   In this letter,  we will  consider {\it  open} strings
(ending  on  D-branes) since  in  string  theory  black holes  can  be
realized as black $p$-brane solutions corresponding to D-branes and it
is  for these  objects one  can provide  a statistical  foundation for
black hole  thermodynamics. 

\
\

Another motivation in searching  open cosmic string solutions in black
hole  backgrounds   originates  from  the   following  considerations.
Consider  in  string theory  a  large  number  of coincident  D-branes
located at $r=0$  in flat space and a  single parallel D-brane located
at $r=d$,  at weak string  coupling $g_s$.  These branes  interact via
long open strings  stretched between them and the  low energy dynamics
of the system can be described  by a suitable gauge theory.  There are
also  closed string  excitations  propagating in  the bulk.  Following
\cite{hor},  let us  try to  understand  what happens  as we  increase
$g_s$.   Specifically, we  are interested  in  the fate  of the  long
string stretched  between $r=0$ and  $r=d$.  As $g_s$  becomes larger,
Newton's  constant $G$ also  increases since  in string  theory $G\sim
g_s^2$  (in units  where $\alpha'=1$).   Thus the  gravitational field
produced by D-branes becomes stronger and at some point it would
be more appropriate to switch  in a curved space description since one
can  no longer  ignore  back reaction  of  the modes  on the  geometry
\cite{hor}.  For a sufficiently large number of coincident branes, one
expects that there would form a black hole (or $p$-brane) which has an
effective Schwarzschild radius proportional  to $GM$, where $M$ is the
mass  of the  coincident  D-branes.   On the  other  hand, the  single
D-brane can be treated as  an hypersurface located outside the horizon
where open stings can end.  If we assume that everything goes smoothly
as  we increase  $g_s$ (which  should be  the case  for an  extreme or
near-extreme configuration),  then in  curved space-time we  should be
able to find  open strings stretched between heavy  D-branes that make
up the hole located inside  the horizon and the single D-brane located
outside.  One   natural  framework   of  investigation  is   to  study
macroscopic cosmic string solutions.  

\
\

In \cite{c8},  an open macroscopic string extending  radially from the
Schwarzschild black hole  horizon was studied. As we  will discuss, the
string of \cite{c8} cannot be interpreted as the open string mentioned
above.  In this paper, we construct different open cosmic string solutions in
black hole  and black  $p$-brane backgrounds which  can be  thought to
stretch between  two D-branes in curved  space-time.  Specifically, we
also try  to determine the fate  of the long  string extending between
the D-branes that make up  the hole and a  test D-brane located
outside  the  horizon.  We  find  that  the  string  splits  into  two
semi-infinite parts  on the  horizon. In some  sense the  horizon acts
like a D-brane  and the strings can be thought  to stretch between the
D-branes  and  the horizon.  However,  the  piece  located inside  the
horizon turns out to be oriented along a time-like direction and gives
rise to tachyonic excitations.  This  shows that this string is unstable and
it will decay by emitting  radiation. We argue that this radiation can
escape  from inside  the horizon  to outside  and we  interpret  it as
Hawking  radiation.  We also  discuss  possible  implications of  this
picture  of  black hole  radiance  on  information  paradox (for  some
attempts  to   solve  the   information  puzzle  see,   for  instance,
\cite{i1}-\cite{is}).

\
\

The  organization  of the  paper  is as  follows.   In  section II  we
construct open cosmic string solutions in Schwarzschild black hole. We
discuss their classical properties and study small fluctuations to the
linearized order. Fixing all world-sheet reparametrization invariance,
one  obtains  free  massless   and  massive  fields  corresponding  to
transverse  fluctuations of the  string. In  section III,  we consider
non-dilatonic black $p$-brane backgrounds and find similar open cosmic
string solutions. Unfortunately, for  $p\not=0$, we find that there is
a  generic  stability  problem   for  the  perturbations  along  brane
directions. In particular, one can  no longer identify the long string
stretched between D-branes that make up the hole and a test
D-brane located outside  the horizon. For $p=0$, however,  there is no
stability problem.  We conclude briefly in section IV.

\section{Open cosmic strings in Schwarzschild black hole}

Consider an open  bosonic string in a vacuum space-time  with the
metric $g_{\mu\nu}$. The motion of this string is determined by the
Polyakov action  
\be \label{pol}
S=\int \,d\ta d\s\,\sqrt{-\g}\,\g^{\al\bt}\,
\partial_{\al}x^{\mu}\,\partial_{\bt}x^{\nu}\,g_{\mu\nu}, 
\ee   
where $\sigma\in[\sigma_i,\sigma_f]$. The world-sheet metric
$\g_{\al\bt}$ is taken to be an independent field. The equations of
motion read 
\bea
T_{\al\bt}&=&\partial_{\al}x^{\mu}\,\partial_{\bt}x^{\nu}\,
g_{\mu\nu}\nonumber\\
&-&\frac{1}{2}\,\g_{\al\bt}\, \g^{ab}\partial_{a}x^{\rho}\partial_{b}x^{\sigma}g_{\rho\sigma}\,=\,0,\label{2}\\
\nabla^2 x^{\mu}\,&+&\,\Gamma^{\mu}_{\rho\sigma}\,\partial_{\al} x^{\rho}\partial_{\bt}x^{\sigma}\g^{\al\bt}\,=\,0,\label{3}
\eea
where $\nabla^{2}$ is the world-sheet Laplacian of  $\g_{\al\bt}$ and
$\Gamma^{\mu}_{\rho\sigma}$ is the Levi-Civita connection of $g_{\mu\nu}$. 
Since we consider open strings, these equations should be 
supplemented by boundary conditions to cancel unwanted surface terms
in the variation of the action.  We will discuss them in a
moment. Using reparametrization invariance we impose the  conformal
gauge where  
\be\label{conf}
\g_{\al\bt}=\eta_{\al\bt}.
\ee 
In this gauge, the momentum conjugate to the  coordinate $x^\mu$ is equal
\be\label{mom}
P_\mu\,=\,\int_{\sigma_i}^{\sigma_f}\,\,\partial_{\tau}x^{\nu}\,\,g_{\mu\nu}\,d\sigma,
\ee
and the total physical length of the string is given by
\be\label{L}
L\,=\,\int_{\sigma_i}^{\sigma_f}\,\,\sqrt{g_{\mu\nu}\,(\partial_\sigma
x^\mu)\,(\partial_\sigma x^\nu)}\,d\sigma.
\ee
Note that thanks to the conformal invariance, (\ref{conf}) still
allows the residual reparametrizations where $\tau+\sigma\to
g(\tau+\sigma)$ and $\tau-\sigma\to h(\tau-\sigma)$. In this paper, we
neglect back reaction of the string on the geometry.

\
\

We consider 4-dimensional Schwarzschild black hole in the
advanced Eddington-Finkelstein coordinates 
\bea
ds^{2}\,&=&-\,f(r)\,dv^2\,+\,2dvdr\,+\,r^2\,d\Omega_2^2\nonumber\\
&+&g_{ij}(x)dx^idx^j,\label{metsch}
\eea
where $f=(1-2M/r)$. Here, we assume a direct product structure with an
internal  space parametrized by $x^i$ which has the metric $g_{ij}$ which 
can be taken to be flat. Our results can easily be generalized to
higher dimensional analogs of the Schwarzschild black hole. 

\
\

We start with the following ansatz 
\be
v=v(\tau,\sigma),\,\,\,\,\,r=r(\tau,\sigma),
\ee
where all other coordinates are set to arbitrary constants. Then
(\ref{2}) implies that
\bea\label{8}
\dot{v}\,r'\,+\,\dot{r}\,v'&=&f\dot{v}\,v',\\
\dot{v}\,\dot{r}\,+\,v'r'&=&\frac{1}{2}\,f\,v'^2\,+\,\frac{1}{2}\,f\,\dot{v}^2,\label{9}
\eea
where the dot and the prime denotes partial differentiation with
respect to $\tau$ and $\sigma$, respectively. A simple solution to
these constraints is given by \footnote{Assuming that $v=\tau-\sigma$,
(\ref{8}) and (\ref{9}) give $\dot{r}-r'=f$. A possible way of
solving this equation is to assume that $r$ depends on $\sigma$,
$\tau$ or $\tau-\sigma$ alone. We choose $r=r(\sigma)$. Choosing
$r=r(\tau)$ is equivalent to interchanging the roles of $\tau$ and
$\sigma$. On the other hand choosing $r=r(\tau-\sigma)$ gives a
collapsed string.\label{f}}  
\be
v=\tau-\sigma,\,\,\,\,\,r=r(\sigma),\label{adv1}
\ee
where 
\be\label{dif1}
\frac{dr}{d\sigma}\,=\,-\,f.
\ee
It is easy to verify that, (\ref{adv1}) also obeys the field equations
(\ref{3}). 

\
\

Eq. (\ref{dif1}) can be solved to obtain an explicit relation between
$r$ and $\sigma$ 
\be\label{soldif}
\sigma\,=\,-r\,-\,2M\ln|2M-r|,
\ee
where we ignore an additive integration constant. As $r\to 2M$ 
$\sigma\to+\infty$. Therefore, the strings cannot cross the event horizon
and are located either inside or outside. It is possible to view them
as one-dimensional semi-infinite throats stretching through the black
hole horizon. The string located inside can be extended through
the singular region $r=0$ for finite $\sigma$. For a generic case,
they are pictured in figure 1. 

\
\

It is worth to note that  the string inside the horizon is oriented
along  a time-like line and   its motion  violates   space-time
causality.  Indeed,  from  (\ref{adv1})  the  induced  metric  on  the
world-sheet can  be found to  be $ds^2=f(-d\tau^2+d\sigma^2)$.  Inside
the horizon $f<0$  and therefore it seems that the  time and the space
coordinates are interchanged. Viewing  $\sigma$ as the time coordinate
alters  the interpretation of  the motion  such that  in figure  1 the
string inside the  horizon should be imagined to  be stretched between
the two  $\tau=\textrm{const.}$ lines and  is moving upwards.  This is
pictured in  figure 2.  The  situation is very similar  to open-closed
string duality where an open string  loop can be expressed in terms of
a  sum over all  closed string  states by  interchanging the  roles of
$\tau$   and  $\sigma$.    In  this   case,  one   dimensional  throat
interpretation should  be understood  along the time  direction.  This
solution   can   also  be   obtained   by   choosing  $r=r(\tau)$   in
(\ref{adv1}). 

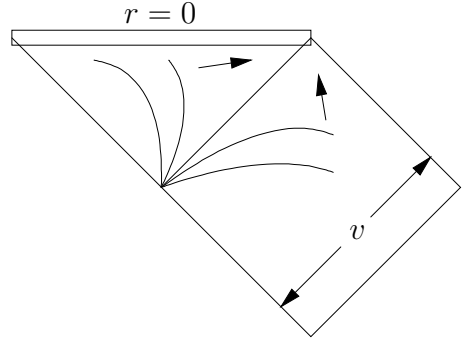
\begin{figure}[htb]
\begin{center}
\setlength{\unitlength}{0.00087489in}
\begingroup\makeatletter\ifx\SetFigFont\undefined%
\gdef\SetFigFont#1#2#3#4#5{%
  \reset@font\fontsize{#1}{#2pt}%
  \fontfamily{#3}\fontseries{#4}\fontshape{#5}%
  \selectfont}%
\fi\endgroup%
{\renewcommand{\dashlinestretch}{30}
\begin{picture}(2724,2052)(0,-10)
\path(12,1857)(1812,1857)(1812,1767)
	(12,1767)(12,1857)
\path(1812,1812)(912,912)
\path(1137,1632)(1452,1677)
\blacken\path(1337.449,1630.331)(1452.000,1677.000)(1328.963,1689.728)(1337.449,1630.331)
\path(12,1812)(1812,12)(2712,912)(1812,1812)
\path(1902,1317)(1857,1587)
\blacken\path(1906.320,1473.565)(1857.000,1587.000)(1847.136,1463.701)(1906.320,1473.565)
\path(1992,552)(1632,192)
\blacken\path(1695.640,298.066)(1632.000,192.000)(1738.066,255.640)(1695.640,298.066)
\path(2172,732)(2532,1092)
\blacken\path(2468.360,985.934)(2532.000,1092.000)(2425.934,1028.360)(2468.360,985.934)
\path(507,1677)(509,1677)(512,1676)
	(518,1675)(526,1673)(538,1671)
	(552,1667)(569,1663)(587,1657)
	(607,1650)(627,1642)(648,1633)
	(669,1621)(691,1607)(712,1591)
	(734,1571)(756,1548)(779,1521)
	(801,1488)(822,1452)(839,1416)
	(855,1380)(867,1343)(878,1307)
	(887,1272)(893,1238)(899,1204)
	(903,1171)(906,1139)(909,1107)
	(911,1077)(912,1047)(912,1019)
	(913,994)(913,971)(913,952)
	(913,937)(912,925)(912,918)
	(912,914)(912,912)
\path(957,1677)(958,1676)(960,1673)
	(964,1668)(969,1661)(975,1652)
	(983,1640)(992,1626)(1001,1611)
	(1009,1594)(1018,1575)(1026,1554)
	(1033,1530)(1039,1504)(1044,1475)
	(1047,1441)(1048,1403)(1047,1362)
	(1044,1323)(1039,1285)(1032,1249)
	(1025,1214)(1016,1181)(1007,1150)
	(998,1120)(988,1091)(977,1064)
	(967,1037)(957,1012)(947,989)
	(938,968)(930,950)(923,936)
	(918,925)(915,918)(913,914)(912,912)
\path(912,912)(913,912)(916,914)
	(922,915)(930,918)(942,922)
	(957,927)(975,933)(996,940)
	(1019,947)(1044,955)(1071,963)
	(1099,971)(1128,979)(1159,987)
	(1190,995)(1223,1003)(1257,1011)
	(1293,1018)(1330,1025)(1370,1031)
	(1411,1037)(1454,1043)(1497,1047)
	(1548,1051)(1595,1053)(1638,1053)
	(1677,1052)(1712,1050)(1744,1047)
	(1773,1044)(1800,1040)(1824,1035)
	(1847,1030)(1868,1025)(1887,1020)
	(1904,1016)(1918,1011)(1929,1008)
	(1937,1005)(1943,1003)(1946,1002)(1947,1002)
\path(912,912)(913,913)(916,915)
	(922,919)(930,925)(942,933)
	(957,944)(975,956)(996,970)
	(1019,986)(1044,1003)(1071,1020)
	(1099,1038)(1128,1056)(1159,1075)
	(1190,1093)(1223,1111)(1257,1129)
	(1293,1146)(1330,1164)(1370,1181)
	(1411,1197)(1454,1213)(1497,1227)
	(1548,1241)(1595,1252)(1638,1260)
	(1677,1265)(1712,1267)(1744,1268)
	(1773,1267)(1800,1264)(1824,1261)
	(1847,1257)(1868,1252)(1887,1247)
	(1904,1242)(1918,1237)(1929,1234)
	(1937,1231)(1943,1229)(1946,1228)(1947,1227)
\put(687,1902){\makebox(0,0)[lb]{\smash{{{\SetFigFont{12}{14.4}{\rmdefault}{\mddefault}{\updefault}$r=0$}}}}}
\put(2042,602){\makebox(0,0)[lb]{\smash{{{\SetFigFont{12}{14.4}{\rmdefault}{\mddefault}{\updefault}$v$}}}}}
\end{picture}}
\end{center}
\caption{The strings in Schwarzschild black hole  located inside and outside the horizon pictured for two different values of $\tau$. The string inside the horizon can be extended through the singular region $r=0$ for finite $\sigma$.}  
\end{figure}

\
\

As pointed out earlier, since we are working with open strings, suitable
boundary conditions should be imposed on the boundary of the
world-sheet. For a string extending through the event horizon
$\sigma\in[\sigma_i,+\infty]$ and thus the only boundary is located at
$\sigma=\sigma_i$. In the generic case, one should also impose a
boundary condition at the maximum value of $\sigma$. Of course we
would like to impose a condition that is obeyed by our solution
(\ref{adv1}). From the variation of the action one obtains the
following surface terms
\be\label{sur}
\delta\,v\,(-f\,\partial_\sigma v\,+\,\partial_\sigma\,r)\,+\,\delta\,r\,(\partial_\sigma\,v),
\ee
which can be canceled by imposing
\be\label{r}
\begin{array}{cc}
\hs{18}\delta\,r=0\\
(\partial_\sigma\,r\,-\,f\,\partial_\sigma \,v)=0
\end{array}
\,\,\,\,\,\textrm{at}\,\,\sigma=\sigma_i,\sigma_f.
\ee
Other coordinates may obey either Dirichlet ($\delta x=0$)  or Neumann
boundary conditions ($\partial_\sigma x=0$).  Note that the first term 
in (\ref{r}) is a condition on fluctuations and the second one 
is satisfied by the string solution (\ref{adv1}). The first condition 
implies that there are two D-branes
located at two different values of $r$, unless $r$ is not on the
horizon (in this case $\sigma$ has an infinite extend and thus there
is no need to impose a boundary condition). Coordinate $r$ is time-like
inside the horizon and in this case it would be more appropriate to
name these as S-branes (i.e. branes having spatial world-volumes). 

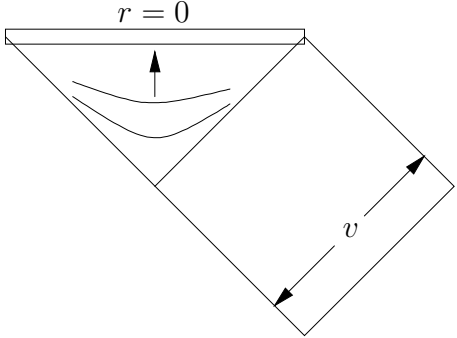
\begin{figure}[htb]
\begin{center}
\setlength{\unitlength}{0.00087489in}
\begingroup\makeatletter\ifx\SetFigFont\undefined%
\gdef\SetFigFont#1#2#3#4#5{%
  \reset@font\fontsize{#1}{#2pt}%
  \fontfamily{#3}\fontseries{#4}\fontshape{#5}%
  \selectfont}%
\fi\endgroup%
{\renewcommand{\dashlinestretch}{30}
\begin{picture}(2724,2052)(0,-10)
\path(12,1857)(1812,1857)(1812,1767)
	(12,1767)(12,1857)
\path(1812,1812)(912,912)
\path(12,1812)(1812,12)(2712,912)(1812,1812)
\path(1992,552)(1632,192)
\blacken\path(1695.640,298.066)(1632.000,192.000)(1738.066,255.640)(1695.640,298.066)
\path(2172,732)(2532,1092)
\blacken\path(2468.360,985.934)(2532.000,1092.000)(2425.934,1028.360)(2468.360,985.934)
\path(912,1452)(912,1722)
\blacken\path(942.000,1602.000)(912.000,1722.000)(882.000,1602.000)(942.000,1602.000)
\path(417,1452)(420,1450)(426,1446)
	(437,1438)(454,1427)(475,1413)
	(502,1396)(532,1376)(565,1356)
	(598,1335)(631,1314)(664,1295)
	(695,1278)(724,1262)(751,1249)
	(777,1237)(800,1228)(823,1220)
	(844,1214)(865,1209)(885,1206)
	(905,1204)(924,1204)(944,1205)
	(964,1208)(985,1212)(1006,1218)
	(1029,1225)(1053,1235)(1078,1246)
	(1106,1258)(1134,1273)(1165,1289)
	(1195,1306)(1226,1324)(1256,1342)
	(1284,1358)(1308,1373)(1328,1386)
	(1344,1395)(1354,1402)(1359,1405)(1362,1407)
\path(417,1542)(420,1541)(428,1538)
	(441,1533)(460,1526)(485,1517)
	(514,1507)(547,1495)(580,1484)
	(614,1473)(647,1462)(679,1452)
	(708,1444)(736,1437)(762,1430)
	(786,1425)(809,1421)(831,1418)
	(853,1416)(875,1414)(894,1414)
	(914,1414)(935,1415)(956,1416)
	(979,1418)(1002,1421)(1028,1425)
	(1055,1429)(1084,1434)(1115,1441)
	(1148,1447)(1181,1454)(1214,1462)
	(1247,1469)(1277,1476)(1304,1483)
	(1325,1488)(1342,1492)(1353,1495)
	(1359,1496)(1362,1497)
\put(687,1902){\makebox(0,0)[lb]{\smash{{{\SetFigFont{12}{14.4}{\rmdefault}{\mddefault}{\updefault}$r=0$}}}}}
\put(2042,602){\makebox(0,0)[lb]{\smash{{{\SetFigFont{12}{14.4}{\rmdefault}{\mddefault}{\updefault}$v$}}}}}
\end{picture}}
\end{center}
\caption{The cosmic string inside the black hole horizon where
$\sigma$ plays the role of world-sheet time. The string leaves the
horizon at $\sigma=-\infty$ and reaches out the singularity at some
finite $\sigma$ forming a semi-infinite throat along time direction.}
\end{figure}

\
\

Viewing $\sigma$ as the time and $\tau$ as the space coordinates,
appropriate  boundary conditions should be imposed at
$\tau=\tau_{i},\tau_{f}$. In this case the surface terms coming from
the variation of the action equal to (\ref{sur}) in which $\partial_\sigma$
is replaced by $\partial_\tau$. These unwanted terms can be canceled
by imposing  
\be\label{r2}
\begin{array}{cc}
\hs{13}\partial_\tau\,r=0\\
f^{-1}\,\delta r\,-\,\delta v=0
\end{array}
\,\,\,\,\,\textrm{at}\,\,\tau=\tau_{i},\tau_{f}.
\ee
The first term in (\ref{r2}) is the Neumann
condition on $r$ and obeyed by our solution (\ref{adv1}) 
and the second one is a Dirichlet condition on the
coordinate defined by $dy=f^{-1}dr - dv$. Inside the horizon $y$ is
a space-like coordinate and thus the second condition implies that
there are two time-like D-branes located at two different values of
$y$. 

\
\

Let us now discuss some of the classical properties of these cosmic strings. 
From (\ref{L}) one can calculate the length as 
\be \label{L2}
L\,=\,\int_i^f\,f^{-1/2}\,dr,
\ee
which converges at $r=2M$. Thus, even though $\sigma$ diverges on the horizon,
the string has a finite proper length. For strings located outside,
(\ref{L2}) diverges linearly as $r\to \infty$ since string stretches
out to spatial infinity. Note that inside the horizon (\ref{L}) gives
an imaginary result consistent with the fact that the string is oriented along
a time-like line. Ignoring the complex number, (\ref{L}) measures the proper
time-like length which converges at $r=0$. From (\ref{mom}) one can
easily calculate non-zero components of momentum vector 
\bea
P_v&=&r_f-r_i\\
P_r&=&\sigma_f-\sigma_i.
\eea
These are conserved quantities. $P_v$ is finite for a string stretched
between two finite values of $r$ and $P_r$ diverges for a string
extending through the horizon. This suggests that, even though the
string has a finite proper length through the horizon, one may need to
introduce a cutoff (a D-brane for the string to end) just before the
string stretches out the horizon.

\
\

On the other hand, inside the event horizon viewing $\tau$ as the
space and $\sigma$ as the time coordinate, (\ref{L}) and (\ref{mom}) should be
modified by replacing $\partial_\sigma$ with $\partial_\tau$ and
$\sigma$ integral by $\tau$ integral. In this case the total length of the
string becomes 
\be 
L\,=\,\sqrt{-f}\,(\tau_f-\tau_i).
\ee
The proper length of the string vanishes on the
horizon. As the string moves through the singularity in the black hole
its length increases. The length finally diverges at $r=0$.
On the other hand the momentum vector can be calculated as 
\bea
P_v&=&f\,(\tau_f-\tau_i),\\
P_r&=&(\tau_f-\tau_i).
\eea
Although $P_r$ is conserved (i.e. $\partial_\sigma P_r=0$) $P_v$ is not
a constant of motion. On the horizon $P_v=0$ and it diverges when the
string hits the singularity at $r=0$.

\
\

In both pictures the space time mass of the string can be calculated
from the center of mass momentum using $m^2=-g^{\mu\nu}P_\mu
P_\nu$. For the strings pictured in figure 1, this is a conserved
quantity, but for the one pictured in figure 2 it is not a constant of
motion.

\
\

After determining these classical properties, let us now study
linearized fluctuations around the solution (\ref{adv1}). This is a
saddle point approximation to the full quantum theory defined by the
non-linear sigma model (\ref{pol}). In this way we can also decide if
the solutions are stable. We define
\bea
v&=&\tau-\sigma \,+\, \delta v,\\
r&=&\bar{r}(\sigma)\,+\,\delta r,
\eea
where $\bar{r}$ is given by (\ref{dif1}).
It turns out that to the linearized order (\ref{2}) involves only $\delta v$ and $\delta r$, and gives two constraints
\bea \label{c11}
\partial_+\,\delta v&=&0,\\
\partial_-\,\delta r\,-\,\frac{1}{2}\,\bar{f}\,\partial_-\delta v\,-\,(\partial_{\bar{r}}\,\bar{f})\,\delta r&=&0,
\label{c22}
\eea
where $\partial_{\pm}=(\partial_\tau\,\pm\,\partial_\sigma)$. On the other hand, $v$ and $r$ components of the field equations (\ref{3}) imply 
\bea
\partial_+\partial_-\,\delta v&=&0,\\
\partial_+\left[\partial_-\,\delta r\,-\,\frac{1}{2}\bar{f}\partial_-\delta v\,-\,(\partial_{\bar{r}}\bar{f})\delta r\right]&=&0.
\eea
Therefore, to this order these field equations are obeyed provided that the constraints (\ref{c11}) and (\ref{c22}) are satisfied.

\
\

It turns out that fixing residual reparametrization invariance one can set $\delta v=\delta r=0$. This can be achieved as follows. From (\ref{c11}) we see that $\delta v$ is a function of $\tau-\sigma$. On the other hand (\ref{c22}) can be rewritten as
\be\label{only}
\partial_-\left(\frac{\delta r}{\bar{f}}-\frac{\delta v}{2}\right)=0.
\ee
One can now perform the following infinitesimal reparametrizations
\bea
\tau-\sigma&\to&\tau-\sigma-\delta v\\
\tau+\sigma&\to&\tau+\sigma+\epsilon(\tau+\sigma)
\eea
where $\epsilon=2\bar{f}^{-1}\delta r-\delta v$. Note that
(\ref{only}) implies that $\epsilon$ is a function of
$\tau+\sigma$. One can easily verify that these transformations set
$\delta v=\delta r=0$. This is very similar to the light cone gauge
where two longitudinal modes were eliminated using residual
reparametrization invariance before quantization. We
expect that it is possible to impose  $\delta v=\delta r=0$ 
systematically order by order in perturbation theory.  

\
\

Following \cite{c8}, it would be interesting to see how
the longitudinal modes decouple in a path integral quantization. As shown by
Polyakov in \cite{pol1}, an arbitrary world-sheet metric can be
parametrized as
\be
h_{\alpha\beta}\,=\,\bar{h}_{\alpha\beta}\delta\lambda\,+\,\bar{\nabla}_{(\alpha}\eta_{\beta)}
\ee
where $\bar{h}_{\alpha\beta}$ is the induced metric and
$\bar{\nabla}$ is the corresponding covariant derivative. Then the
functional integration measure over the metrics can be decomposed as
\be\label{fp}
{\cal D}h_{\alpha\beta}\,=\,{\cal D}\lambda\,{\cal D}\eta_\alpha\,\det[L^{\dagger}L]^{1/2},
\ee
where $L^{\dagger}L$ is the following operator acting on one-forms
\be\label{LL}
(L^{\dagger}L)\eta_{\alpha}=(\bar{h}_{\alpha\beta}\bar{\nabla}^{2}+[\bar{\nabla}_{\beta},\bar{\nabla}_{\alpha}])\eta^{\beta}.
\ee
The integral over $\lambda$ decouples since
the background obeys $\beta$-function equations to first order in
$\alpha'$ (since the black hole is Ricci flat).
We now show that the Fadeev-Popov determinant involving the
operator $L$ cancels against the determinant arising from the
integration of longitudinal modes.

\
\

This can easily be seen by a normal coordinate expansion of the action
(\ref{pol}) around the classical solution (\ref{adv1}). To quadratic
order the terms involving the longitudinal modes can be written as 
\bea
S&=&\int d\tau d\sigma\,\, \eta_{ab}\,\eta^{\alpha\beta} \,D_\alpha
\eta^a\,D_\beta\eta^b\nonumber\\
&+&
\eta^{\alpha\beta}R_{\mu ab\nu}\,\eta^a\,\eta^b\,\partial_\alpha\bar{\eta}^\mu\partial_\beta\bar{\eta}^\nu, \label{normal}
\eea
where 
\be\label{D}
D_\alpha\eta^a\,=\,\partial_\alpha\eta^a\,+\,w_{\mu}^{ab}(\partial_{\alpha}\bar{\eta}^{\mu})\eta_b,
\ee
$R_{\mu\nu\rho\sigma}$ and $w_{\mu}^{ab}$ are the curvature
tensor and spin connection of the space-time metric, $\bar{\eta}^\mu$ is the
solution (\ref{adv1}) and $\eta^a$ represents $(\delta v,\delta r)$ in
normal coordinate form. It would be more convenient to  change the
coordinates using $du=dv-2dr/f$ to recast $\bar{\eta}^\mu$ in a much
simpler form. Note that $u$ is the retarded null coordinate. 
In $(u,v)$ coordinates, the solution (\ref{adv1}) 
becomes $u=\tau+\sigma$ and $v=\tau-\sigma$. Thus $(u,v)$ plane can be
identified with the $(\tau,\sigma)$ plane which implies that
$D_\alpha=\bar{\nabla}_\alpha$ and the last term in (\ref{normal})
becomes $-\bar{R}_{ab}\eta^a\eta^b$ where $\bar{R}_{ab}$ is the Ricci
tensor of $\bar{h}_{\alpha\beta}$. Integrating by parts, one exactly
obtains the operator $L^\dagger L$ in (\ref{LL}) contracted by two
$\eta^a$ from left and right. The functional integral over $\eta^a$
then gives $\det[L^\dagger L]^{-1/2}$ exactly canceling the
Fadeev-Popov determinant in (\ref{fp}). 

\
\

This shows that the longitudinal modes decouple and   
we are left with the transverse fluctuations of the string along
internal directions parametrized by $\delta x$ and the angular
directions on $\Omega_2$ parametrized by $\delta \theta$. In a normal
coordinate expansion, (\ref{3}) gives field equations for these
perturbations which read 
\bea
\partial^\alpha\partial_\alpha\,\delta \theta\,-\, 2\,M\,(\bar{r})^{-3}\,\bar{f}\,\delta \theta&=&0\label{f1}\\
\partial^\alpha\partial_\alpha\,\delta x&=&0 \label{f2}.
\eea
Thus the perturbations along internal directions, $\delta x$, obey free
massless scalar equation on the world-sheet. On the other hand, there
is a mass term for angular perturbations, $\delta\theta$, on $\Omega_2$.
Not surprisingly  (\ref{f1}) is ill  defined at $\bar{r}=0$  i.e. when
the string hits the singularity.

\
\

For the string outside the horizon the ``mass squared'' 
term in (\ref{f1}) has the correct
sign. Thus small fluctuations around this solution does not run away
which shows that the configuration is stable. The mass depends on the
radial position of the string which vanish on the horizon and at the
spatial infinity  at $r=\infty$. However, contrary to the macroscopic
string considered in \cite{c8}, there is a time translation symmetry
on the world-sheet and thus there is a globally well defined notion of
a particle. Thus there is no particle creation by the gravitational
field and Hawking radiation cannot be realized along this string.

\
\

For the string inside the horizon pictured in figure 2 (where the
roles of the $\tau$ and $\sigma$ are interchanged as world-sheet space 
and time) the ``mass squared''  term in (\ref{f1}) has also the
correct sign. Thus, this solution is also stable under
small perturbations. (Note that in this case $\bar{f}<0$.) 
However, the mass term depends on world-sheet time
$\sigma$, therefore this is not a static configuration. We expect that
there will be particle creation on the world-sheet 
by gravitational field. 
One can define the Fock vacuum when the string leaves the horizon at
$\sigma=-\infty$. Since creation and annihilation operators are time
dependent, this vacuum state will evolve to some many particle
state. However this radiation is confined in the horizon and again
black hole radiance cannot be realized along such a string.   

\
\

For the  string inside  the horizon pictured  in figure 1,  the ``mass
squared'' term has the {\it wrong} sign which indicates that there are
tachyonic excitations. This is  a highly unstable configuration and one
expects that  it will  decay to  a stable ground  state by  giving out
radiation. However, world-sheet is embedded in space-time so that some
of  the  perturbations  move  on classically  forbidden  paths.   More
precisely,  the right  moving and  left moving  massless  modes follow
$\tau-\sigma=\textrm{const.}$  and $\tau+\sigma=\textrm{const.}$ lines
in  the world-sheet,  respectively.   By the  map (\ref{adv1}),  these
modes follow $v=\textrm{const.}$  and $u=\textrm{const.}$ lines, where
$u$  is the  retarded  null coordinate  define  below (\ref{D}).   The
directions for these  modes are pictured in figure  3. The left moving
modes follow constant $u$ lines and propagate on causally well defined
paths.   They  hit  the  singularity  at  $r=0$  in  a  finite  proper
world-sheet time. On the other  hand, the right moving modes propagate
through  the  horizon  on  classically  forbidden  $v=\textrm{const.}$
paths.  Since $\sigma$  diverges on the horizon these  modes can never
reach  the horizon,  but they  can carry  radiation from  $r=0$  to an
arbitrarily  close distance  to the  horizon on  a constant  $v$ line.
Assuming  small  fluctuations  of   the  light  cone  on  the  horizon
(corresponding to  small perturbations of the metric),  we expect that
these modes can escape out  to infinity.  Therefore, these strings can
carry  radiation from  inside  the  horizon to  the  horizon and  that
radiation can escape out to infinity in the form of Hawking radiation.
The  thermal   nature  of  the   radiation  can  be  related   to  the
thermodynamics of  the massless modes  moving on the  world-sheet (for
instance as discussed in \cite{mat}).   It is interesting to note that
in  this   scenario  we  are   not  dealing  with  the   full  quantum
gravitational  effects.  Namely,  we  still cannot  exactly solve  the
theory defined  by (\ref{pol}). Thus the singularity  problem at $r=0$
still   survives. However, 
information puzzle can be resolved since the
radiation which escape out to infinity can carry information about the
collapsed matter at $r=0$. The  information  transfer   is  achieved
in a semiclassical context by the help of the radiation  tunneling
through the horizon along classically forbidden paths.

\begin{figure}[htb]
\begin{center}
\setlength{\unitlength}{0.00087489in}
\begingroup\makeatletter\ifx\SetFigFont\undefined%
\gdef\SetFigFont#1#2#3#4#5{%
  \reset@font\fontsize{#1}{#2pt}%
  \fontfamily{#3}\fontseries{#4}\fontshape{#5}%
  \selectfont}%
\fi\endgroup%
{\renewcommand{\dashlinestretch}{30}
\begin{picture}(2229,1962)(0,-10)
\path(12,1722)(1812,1722)(1812,1632)
	(12,1632)(12,1722)
\path(12,1677)(1677,12)
\path(1812,1677)(912,777)
\path(1812,1677)(2217,1272)
\path(1722,687)(1362,327)
\blacken\path(1425.640,433.066)(1362.000,327.000)(1468.066,390.640)(1425.640,433.066)
\path(1902,867)(2217,1182)
\blacken\path(2153.360,1075.934)(2217.000,1182.000)(2110.934,1118.360)(2153.360,1075.934)
\path(642,1272)(957,957)
\blacken\path(850.934,1020.640)(957.000,957.000)(893.360,1063.066)(850.934,1020.640)
\path(642,1272)(957,1587)
\blacken\path(893.360,1480.934)(957.000,1587.000)(850.934,1523.360)(893.360,1480.934)
\put(642,1812){\makebox(0,0)[lb]{\smash{{{\SetFigFont{12}{14.4}{\rmdefault}{\mddefault}{\updefault}$r=0$}}}}}
\put(1362,1092){\makebox(0,0)[lb]{\smash{{{\SetFigFont{12}{14.4}{\rmdefault}{\mddefault}{\updefault}$r=2M$}}}}}
\put(1812,732){\makebox(0,0)[lb]{\smash{{{\SetFigFont{12}{14.4}{\rmdefault}{\mddefault}{\updefault}$v$}}}}}
\end{picture}}
\end{center}
\caption{The directions for the right and the left moving modes in space-time, which follow constant $v$ and $u$ lines, respectively.}
\end{figure}
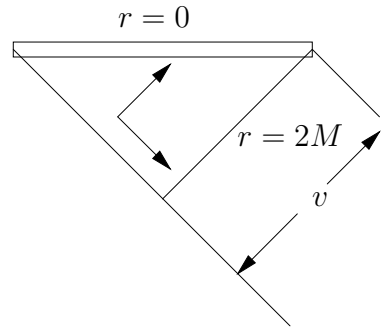

\
\

At this  point one  may simply suggest  to discard this  cosmic string
solution since it gives rise  to tachyonic excitations. However, we now
argue that such a configuration naturally exists in the formation of a
black hole. Let us recall from our discussion in the introduction that
at weak  string coupling there  exist an open string  stretched between
two D-branes i.e. between $r=0$ and $r=d$ ($d>2M$), and as we increase
the coupling, the back reaction  of the heavy D-branes on the geometry
should  be  included in  the  description.  This  can be  achieved  by
switching on the  function $f$ in the metric  (\ref{metsch}). Note that
when $f=0$ we  have the flat space.  This gives  a geometric flow from
flat space to the curved  black hole background.  In this flow the
open string stretched between the two D-branes can now be obtained from
(\ref{adv1})  where $f$ is  changing with  the string  coupling. There
would form  a horizon when $f$  has a zero. Before  this happens, open
string  is oriented  along  a  space-like direction  and  there is  no
violation of causality and no tachyonic excitations. The string extends
smoothly from $r=0$ to $r=d$.  As $f$ tends to zero at $r=2M$, (\ref{dif1})
implies  that $\sigma$  starts to  diverge and  the string  inside the
horizon  tends  to be  stretched  more  and  more through  a  time-like
direction. When $f$ acquires the zero, the string splits into two semi-infinite
parts and the one located inside the horizon now finds itself to be
oriented  along a time-like direction. As  discussed above  this is  not a
stable configuration and this string decays by giving out radiation.

\
\

In \cite{c8} a different mechanism for black hole evaporation along
macroscopic strings were proposed where the Hawking radiation was
interpreted as the thermal bath of the string modes. 
For completeness let us also
discuss the cosmic strings considered in \cite{c8}. In Kruskal
coordinates the metric of the Schwarzschild black hole becomes  
\be\label{met1}
ds^{2}\,=\,F(r)\,dUdV\,+\,r^2\,d\Omega_2^2\,+\,dx^idx^i,
\ee 
where 
\be
F(r)\,=\,\frac{16M^2}{r}\,e^{-r/2M}
\ee
and $r$ is determined by the relation 
\be\label{uv}
UV\,=\,(r-2M)\,e^{r/2M}.
\ee
On this background, one can easily verify that there is a cosmic
string solution to (\ref{2}) and (\ref{3}) which is given by
\bea
V\,=\,\tau+\sigma,\,\,\, U\,=\,-\tau+\sigma, \label{sol1}
\eea
where all other coordinates are set to arbitrary constants.
In fact, one can find a more general solution 
$U=g(\tau+\sigma)$ and $V=g(-\tau+\sigma)$, where $g$ is an arbitrary
function. However, by a coordinate transformation on the world-sheet,
that preserves the conformal gauge (\ref{conf}), one can set $g$ to identity. 

\
\

In \cite{c8}, the authors considered an infinitely long open
string. However, it is possible to obtain strings having finite width 
by imposing suitable boundary conditions at $\sigma=\sigma_i$ and
$\sigma=\sigma_f$ to cancel unwanted surface terms in the
variation. A convenient choice is
\be\label{c111}
\begin{array}{cc}
\delta(U+V)\,=\,0\\
\partial_{\sigma}(V-U)\,=\,0 
\end{array}
\,\,\,\,\textrm{at}\,\,\sigma=\sigma_i,\sigma_f,
\ee
where the other coordinates may obey Dirichlet or Neumann boundary
conditions. The first term in (\ref{c111}) is a condition on the
fluctuations and the second term is satisfied by the cosmic string
solution (\ref{sol1}). On the other hand, the first condition 
 implies that there are two D-branes located at $U+V=2\sigma_i$ and
$U+V=2\sigma_f$.   

\
\

Note that $t=V-U$ and $x=V+U$ are time-like and
space-like coordinates, respectively. Each point on the string
follows the non-geodesic time-like curve parametrized by $t$.
It starts from the past singularity at $\tau=-\sqrt{\sigma^2+2M}$ 
and reaches out the future one at $\tau=+\sqrt{\sigma^2+2M}$.
Depending on $\sigma_i$  and $\sigma_f$, 
the string extends across the past and the future event horizons 
in the maximally extended black hole background. 
For a generic case, the motion of the string is pictured in the
Penrose diagram in figure 4. For an infinitely long string $(t,x)$ plane
can be identified with the world-sheet spanned by $(\tau,\sigma)$
coordinates. 

\
\

\begin{figure}[htb]
\begin{center}
\setlength{\unitlength}{0.00087489in}
\begingroup\makeatletter\ifx\SetFigFont\undefined%
\gdef\SetFigFont#1#2#3#4#5{%
  \reset@font\fontsize{#1}{#2pt}%
  \fontfamily{#3}\fontseries{#4}\fontshape{#5}%
  \selectfont}%
\fi\endgroup%
{\renewcommand{\dashlinestretch}{30}
\begin{picture}(2724,1770)(0,-10)
\path(687,135)(2037,135)(2037,225)
	(687,225)(687,135)
\path(687,1530)(12,855)(687,180)
\path(822,495)(2037,495)
\path(1452,630)(1452,1215)
\blacken\path(1482.000,1095.000)(1452.000,1215.000)(1422.000,1095.000)(1482.000,1095.000)
\drawline(1182,135)(1182,135)
\path(2037,1530)(2712,855)(2037,180)
\path(687,1530)(2037,180)
\path(2037,1530)(687,180)
\path(687,1575)(2037,1575)(2037,1485)
	(687,1485)(687,1575)
\put(1200,1620){\makebox(0,0)[lb]{\smash{{{\SetFigFont{12}{14.4}{\rmdefault}{\mddefault}{\updefault}$r=0$}}}}}
\put(1200,-50){\makebox(0,0)[lb]{\smash{{{\SetFigFont{12}{14.4}{\rmdefault}{\mddefault}{\updefault}$r=0$}}}}}
\end{picture}}
\end{center}
\caption{The maximally extended black hole and the cosmic string of \cite{c8}.}
\end{figure}
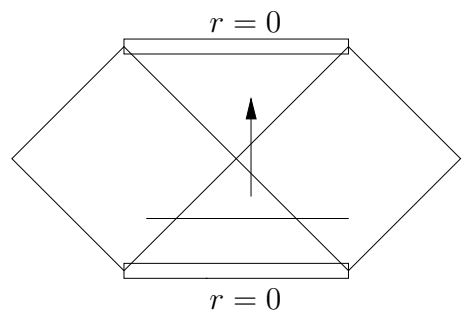

From (\ref{mom}), the only non-zero components of the momentum vector
become $P_v=-P_u=P$, where $P$ is given by 
\be\label{P}
P\,=\,\frac{1}{2}\,\int_{\sigma_i}^{\sigma_f}\,F\,d\sigma.
\ee
One can easily verify that $\partial_\tau P\not=0$. $P$ diverges when
the string hits the past or the future singularities since $F$ blows
up, but otherwise it is finite. In $(t,x)$ coordinates, the momentum
vector becomes, $P_t=P$ and $P_x=0$. Therefore (\ref{P}) measures the
total energy of the string with respect to $t$. Since
$\partial_t$ is not a Killing vector of (\ref{met1}), it is not
surprising that the energy is not conserved. On the other hand, the total
length of the string is given by the integral (\ref{P}) in which $F$
is replaced by $\sqrt{F}$. The length is not a constant of motion and
is finite unless the string hits the singularities.

\
\

Looking   at  small   perturbations  around   this   classical  string
background, one  can show that  it is possible  to gauge away  the two
longitudinal modes using residual reparametrization invariance.  Thus,
one ends  up with  the transverse fluctuations  of the  string obeying
\cite{c8} 
\bea
\partial^\alpha\partial_\alpha\delta\theta\,-\,\bar{r}^{-1}(\partial^\alpha\partial_\alpha\bar{r})\delta\theta&=&0,\label{t}\\
\partial^\alpha\partial_\alpha\,\delta x&=&0,\label{x} 
\eea  where
$\bar{r}$ is the background value  of $r$ which can be determined from
(\ref{uv}). The ``mass squared''  term in (\ref{t})
has the correct sign in the maximally extended black hole and thus the
classical string  is stable under  small perturbations. The  mass term
vanishes on the horizon and at the spatial infinity. Since this is not
a  static  configuration, one  expects  particle  creation effects  by
gravitational field.  Indeed looking at the part of the string that is
located  outside the  horizon it  is  possible to  recover black  hole
radiance by  carefully defining observer dependent  ``in'' and ``out''
vacua  \cite{c8}.  The  relation  between this  approach  and the  one
proposed  in  this letter  is  not clear  to  us.  However, note  that
(\ref{sol1}) cannot represent the string stretched between the D-branes
that make up the hole and a test D-brane located outside the horizon
since  it does  not intersect  $r=0$  (where the  D-branes inside  the
horizon are located) for some time during its motion.  

\section{Open Cosmic Strings in black $p$-brane backgrounds}

In this section, we consider non-dilatonic black $p$-brane backgrounds
of \cite{hor2}. They can be  obtained as magnetically charged
solutions of a theory involving only the metric and a $q$-form field
$F_q$ which has the following simple action 
\be
S=\int\,d^{d}x\,\sqrt{g}\,(R-F_q^2),
\ee
where $q=d-p-2$. The metric can be written as 
\bea
ds^2&=&-\Delta_+\Delta_-^b\,dt^2\,+\,(\Delta_+\Delta_-)^{-1}\,dr^2\nonumber\\  
&+&\Delta_-^{b+1}(d\vec{y}.d\vec{y}) \,+\,r^2d\Omega_{q}^2, \label{met11}
\eea
where 
\be
\Delta_{\pm}=1-\left[\frac{r_\pm}{r}\right]^{q-1},
\ee
and $b=(1-p)/(1+p)$. The  non-extremal solutions have $r_+>r_-$, where
there   are  regular   event   horizons  at   $r=r_+$  and   curvature
singularities  at $r=r_-$.  The  solutions cannot  be extended  beyond
$r<r_-$ when $b$ is not  an integer (i.e. unless $p=0,1$). The extreme
solutions  have $r_+=r_-$  and  $r=r_+$ is  a  regular event  horizon.
These are known to be  stable when embedded in a suitable supergravity
since  they preserve  some supersymmetry.   Thus there  is  no Hawking
radiation on  extreme backgrounds. For  appropriate values of  $d$ and
$p$, (\ref{met11}) includes non-dilatonic extended object solutions in
string/M theory like the the  membrane and the five-brane solutions of
$D=11$ supergravity  and the self-dual three-brane of  $D=10$ type IIB
supergravity.  

\
\

It is possible to introduce an advanced null coordinate defined by
\be
dv\,=\,dt\,+\,\Delta_+\Delta_-^{-(b+1)/2}\,dr
\ee
so that there is no apparent coordinate singularity on the horizon at
$r=r_+$. The metric (\ref{met11}) then becomes
\bea
ds^2&=&-\Delta_+\Delta_-^b\,dv^2\,+\,2\Delta_-^{(b-1)/2}dv\,dr\nonumber\\
&+&\Delta_-^{b+1}(d\vec{y}.d\vec{y})\,+\,r^2d\Omega_{q}^2.\label{met22}
\eea
In this coordinate system unless $p=0$ (i.e. $b=1$) the region $r<r_-$
is not included in space-time. Similarly, in the extreme limit, $r<r_+$
region is only well defined for $b=1$

\
\

On these backgrounds the string propagation is still described by the
action (\ref{pol}) since there is no dilaton and these are
magnetically charged solutions. One can then easily verify that
there is a solution to (\ref{2}) and (\ref{3}) which is given by
\be
v=\tau-\sigma,\,\,\,\,\,r=r(\sigma),\label{adv3}
\ee
where 
\be\label{dif3}
\frac{dr}{d\sigma}\,=\,-\,\Delta_+\Delta_-^{(b+1)/2}.
\ee
It is possible to obtain an explicit relation between $r$ and $\sigma$
by integrating this differential equation. As in the Schwarzschild black
hole, $\sigma$ diverges at $r=r_+$ and thus the strings are located either
inside or outside the horizon. The string inside the horizon 
can be extended through
$r=r_-$ for finite $\sigma$ when $b\not=1$. When $b=1$ (i.e. $p=0$),
$\sigma$ also diverges at $r=r_-$. There are now
three different pieces which are located in between $(0,r_-)$,
$(r_-,r_+)$ and outside the horizon. The one stretched between
$(r_-,r_+)$ has $\sigma\in(-\infty,+\infty)$. 

\
\

To be able to determine the orientations of the strings in space-time,
the induced metric on the world-sheet can be found to be
\be\label{ind22}
ds_{ind}^2\,=\,\Delta_-^b\,\Delta_+\,(-d\tau^2\,+\,d\sigma^2).
\ee
The strings located outside the horizons are always oriented along
space-like directions. In non-extreme backgrounds, when the string is
located in between $(r_-,r_+)$, the world-sheet space and time coordinates are
interchanged. This string is oriented along a time-like direction, and
one expects that this is an unstable configuration. When $b=1$, the string
located in the region $(0,r_-)$ is oriented along a space-like
direction. In the extreme limit, for $b=1$, the string located inside
the horizon also oriented along a space-like direction.

\
\

In the variation of the action, to  cancel the unwanted surface terms
one can impose
\bea
\delta r&=&0,\\
\partial_\sigma r-\Delta^{\frac{(b+1)}{2}}_-\Delta_+\partial_\sigma
v&=&0,
\eea
where all other coordinates may obey either Dirichlet or Neumann
conditions. The first condition implies that the cosmic string
stretches between two D-branes located at different values of $r$, and
the second condition is obeyed by the cosmic string solution
(\ref{adv3}).
 
\
\

We now show that using residual reparametrization invariance one can
set $\delta v=\delta r=0$. To see this, expanding (\ref{3}) to the
linearized order in perturbations, we obtain
\bea
\partial_+\delta v=0\label{pv}\\
\Delta_-^{\frac{(b-1)}{2}}\partial_-\delta
r\,-\,\frac{b+1}{2}\Delta_-^{(b-1)}\Delta_+\Delta_-'\,\delta
r\nonumber\\
-\,\Delta_-^b\Delta_+'\,\delta r -
\frac{1}{2}\Delta_-^b\Delta_+\,\partial_-\delta v =0 \label{pr},
\eea
where $'$ denotes differentiation with respect to $r$ and all functions 
are evaluated on the background cosmic string solution
(\ref{adv3}). Eq. (\ref{pv}) implies that $\delta v$ is a function of
$\tau-\sigma$. One can then show that the following residual
reparametrizations set $\delta v=\delta r=0$; 
\bea
\tau-\sigma&\to&\tau-\sigma-\delta v\\
\tau+\sigma&\to&\tau+\sigma+\epsilon(\tau+\sigma),
\eea
where $\epsilon=2\Delta_-^{-(b+1)/2}\Delta_+^{-1}\delta
r-\delta v$. By (\ref{pr}), we have $\partial_-\epsilon =0$.

\
\

Therefore, as in the black hole case, longitudinal modes can be gauged
away. On the other hand by a normal coordinate expansion, the
transverse fluctuations can be determined to obey
\bea
\partial_\alpha\partial^\alpha\delta\theta\,-\,m_\theta^2\,\delta\theta&=&0,\\
\partial_\alpha\partial^\alpha\delta y\,-\,m_y^2\,\delta y&=&0,
\eea
where $\delta\theta$ and $\delta y$ represent perturbations along the
transverse $q$-sphere and the brane directions in (\ref{met22}) and the mass
squared terms are given by
\bea
m_\theta^2&=&\frac{\Delta_+\Delta^b_-}{r}\left[\Delta_+\Delta_-\right]',\\
m_y^2&=&\frac{(b+1)}{2}\Delta_+\left[\Delta_-^b\Delta_+\Delta_-'\right]',
\eea
where $'$ denotes differentiation with respect $r$. For the strings
located outside the horizon, i.e. when $r>r_+$, $m_\theta^2>0$ and
$m_y^2$ can be zero, positive or negative depending on $r$. When $r_-<r<r_+$
both $m_\theta^2$ and $m_y^2$ can be positive or negative. And finally
when $b=1$ in the region $r<r_-$, $m_\theta^2<0$ and $m_y^2$ can be
positive, negative or zero. 

\
\

Therefore,  as in the  Schwarzschild black  hole, the  strings located
inside  the horizons are  always unstable.  However, the  ones located
outside are stable for certain  values $r$, i.e. when the D-branes are
placed  in  certain  regions.  The  main  unstability  is  related  to
perturbations   along   brane   directions.  Therefore,   unlike   the
Schwarzschild  black hole, these  strings cannot be  interpreted as
the  long open
strings  stretched  between  the  D-branes  that  make  up  the  black
$p$-brane and a test D-brane located outside the horizon unless $p=0$.
It  seems one  should  try different  ansatzs  (perhaps a  non-trivial
$\sigma$  dependence  of  $y$  coordinates).  We  will  consider  this
possibility  in  a future  work.  However,  one  can still  insist  on
realizing Hawking radiation as a  result of decaying of an open string
stretched between  $r=r_-$ and  $r=r_+$.  As in  the black  hole case,
since this string is oriented along a time-like line by (\ref{ind22}),
one can show  that the right moving massless  modes on the world-sheet
follow causally forbidden  paths, cross the horizon and  escape out to
infinity.  As a consistency check on this interpretation, we note that
in the  extreme solution for $b=1$  i.e. when $p=0$  (in other extreme
cases $r<r_+$ region is not included in space-time) the string located
inside  the   horizon  is  oriented  along   a  space-like  direction.
Therefore, when the extreme limit is reached then the radiation can no
longer escape out to infinity and is confined in the horizon.

\section{Conclusions}

In  this   paper,  we  found   classical  open  string   solutions  in
Schwarzschild   black   hole   and   non-dilatonic   black   $p$-brane
backgrounds.   We   also  studied  small   fluctuations  around  these
macroscopic configurations.  Classical and semiclassical closed string
propagation have  been studied in the literature  before.  We extended
these considerations  to open strings.  Working with  open strings, we
had  to impose  boundary conditions  and  it turns  out that  suitable
conditions  exist such  that open  strings can  be thought  to stretch
between  D-branes or  between a  D-brane and  the horizon.   While the
string solutions  located outside the  horizons turn out to  be stable
against small perturbations in  the Schwarzschild black hole and black
0-branes,  there  is  a  generic  stability  problem  in  other  black
$p$-brane backgrounds.  On the other hand, strings  located inside the
horizons are always unstable. We argued that such strings decay by giving out
radiation and that radiation can escape out to infinity in the form of
Hawking radiation.  

\
\

\begin{figure}[htb]
\begin{center}
\setlength{\unitlength}{0.00087489in}
\begingroup\makeatletter\ifx\SetFigFont\undefined%
\gdef\SetFigFont#1#2#3#4#5{%
  \reset@font\fontsize{#1}{#2pt}%
  \fontfamily{#3}\fontseries{#4}\fontshape{#5}%
  \selectfont}%
\fi\endgroup%
{\renewcommand{\dashlinestretch}{30}
\begin{picture}(2429,2016)(0,-10)
\path(1812,1812)(912,912)
\path(1137,1632)(1452,1677)
\blacken\path(1337.449,1630.331)(1452.000,1677.000)(1328.963,1689.728)(1337.449,1630.331)
\path(1812,1812)(12,1812)(1812,12)(1812,1812)
\path(1632,1182)(1677,1452)
\blacken\path(1686.864,1328.701)(1677.000,1452.000)(1627.680,1338.565)(1686.864,1328.701)
\path(507,1677)(509,1677)(512,1676)
	(518,1675)(526,1673)(538,1671)
	(552,1667)(569,1663)(587,1657)
	(607,1650)(627,1642)(648,1633)
	(669,1621)(691,1607)(712,1591)
	(734,1571)(756,1548)(779,1521)
	(801,1488)(822,1452)(839,1416)
	(855,1380)(867,1343)(878,1307)
	(887,1272)(893,1238)(899,1204)
	(903,1171)(906,1139)(909,1107)
	(911,1077)(912,1047)(912,1019)
	(913,994)(913,971)(913,952)
	(913,937)(912,925)(912,918)
	(912,914)(912,912)
\path(957,1677)(958,1676)(960,1673)
	(964,1668)(969,1661)(975,1652)
	(983,1640)(992,1626)(1001,1611)
	(1009,1594)(1018,1575)(1026,1554)
	(1033,1530)(1039,1504)(1044,1475)
	(1047,1441)(1048,1403)(1047,1362)
	(1044,1323)(1039,1285)(1032,1249)
	(1025,1214)(1016,1181)(1007,1150)
	(998,1120)(988,1091)(977,1064)
	(967,1037)(957,1012)(947,989)
	(938,968)(930,950)(923,936)
	(918,925)(915,918)(913,914)(912,912)
\path(912,912)(914,912)(917,913)
	(923,914)(933,916)(945,918)
	(961,921)(979,924)(1000,928)
	(1022,932)(1045,935)(1070,939)
	(1095,942)(1122,946)(1150,949)
	(1180,951)(1212,954)(1246,955)
	(1281,957)(1317,957)(1361,956)
	(1401,955)(1436,952)(1467,949)
	(1494,945)(1519,941)(1541,936)
	(1562,932)(1580,927)(1596,923)
	(1609,919)(1619,916)(1626,914)
	(1630,913)(1632,912)
\path(912,912)(914,913)(917,914)
	(923,917)(933,920)(945,925)
	(961,931)(979,938)(1000,946)
	(1022,954)(1045,963)(1070,972)
	(1095,981)(1122,990)(1150,999)
	(1180,1009)(1212,1018)(1246,1028)
	(1281,1038)(1317,1047)(1361,1057)
	(1401,1066)(1436,1073)(1467,1078)
	(1494,1082)(1519,1085)(1541,1088)
	(1562,1089)(1580,1090)(1596,1091)
	(1609,1092)(1619,1092)(1626,1092)
	(1630,1092)(1632,1092)
\put(642,1857){\makebox(0,0)[lb]{\smash{{{\SetFigFont{12}{14.4}{\rmdefault}{\mddefault}{\updefault}$r=\infty$}}}}}
\put(597,777){\makebox(0,0)[lb]{\smash{{{\SetFigFont{12}{14.4}{\rmdefault}{\mddefault}{\updefault}$v=-\infty$}}}}}
\put(1722,1812){\makebox(0,0)[lb]{\smash{{{\SetFigFont{12}{14.4}{\rmdefault}{\mddefault}{\updefault}$v=+\infty$}}}}}
\put(1857,732){\makebox(0,0)[lb]{\smash{{{\SetFigFont{12}{14.4}{\rmdefault}{\mddefault}{\updefault}$r=0$}}}}}
\end{picture}}
\end{center}
\caption{The strings in de Sitter space located inside and outside the
cosmological horizon pictured for two different values of world-sheet time.} 
\end{figure}
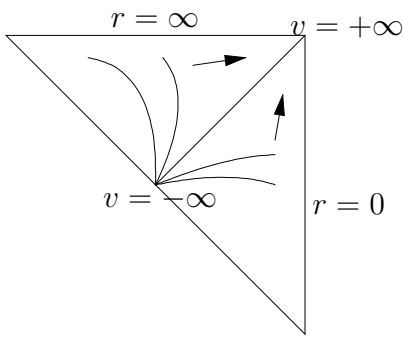

As discussed in section II, our formulation of  black hole radiance  
might offer a resolution  of information paradox.  The strings located
inside the horizons are embedded in space-time so that the world-sheet
times run along spatial directions.  In that respect, they are similar
to  instantons of  field  theories.  Moreover,  like instantons,  they
allow  classically  forbidden  transitions;  the strings  connect  the
singular region  to the  horizon and carry  out energy  on classically
forbidden paths. Note that all these happen inside the horizon and are
not visible  to an asymptotic observer.   

\
\

One can try to combine
ideas  presented in  this paper  with  the others.   For instance,  in
\cite{is},  Mathur argued  that  the  bound states  of  branes have  a
non-zero size  which grows  with the number  of branes. In  this case,
non-local effects induced by instantonic strings may operate in a much
smaller scale. It would also be interesting to incorporate the idea of
the stretched horizon \cite{i10} in this picture.

\
\

It is possible to  extend  these  considerations  in  many  different
directions.  For instance, one can  try to construct cosmic strings in
other black  $p$-brane solutions  corresponding to D-branes  where the
dilaton  have non-trivial  vacuum  expectation values.   One can  also
investigate  what  would  happen  when  the string  couples  to  other
background fields like NS-NS two form potential.  In addition, one can
consider  superstrings  to  see   if  they  can  carry  out  fermionic
information. In all cases, it would be interesting to redrive
Hawking's classical result  on black hole radiance.

\
\

Finally, let us note that it is possible to construct similar cosmic
string solutions in de Sitter space. Indeed in the advanced
Eddington-Finkelstein coordinates the de Sitter metric can be written as  
\be
ds^{2}\,=\,f(r)\,dv^2\,-\,2dvdr\,+\,r^2\,d\Omega_2^2
\ee
where $f=(r^2-1)$, which is very similar to the black hole metric
(\ref{metsch}). In this coordinate system, the string solution is given
by (\ref{adv1}) where $f$ is replaced by $r^2-1$ (see figure
5). These strings have similar properties with the ones that are
propagating in black holes. For instance, they also strech through the
cosmological horizon and it is possible to see that 
inside the horizon they are located along time-like
directions and  give rise to tachyonic excitations.  
However, let us point out that currently it is not
known how to embed de Sitter space in string theory. 
One technical issue related to this problem is that it seems difficult
to impose the conformal gauge (\ref{conf}) since it is not possible to
satisfy the world-sheet $\beta$-function equations. Therefore, our
understanding of open cosmic strings propagating in de Sitter  space
is not complete.

\end{document}